\documentclass[11pt]{article}
\usepackage{epsfig}
\usepackage{amsfonts}
\usepackage{amsmath}
\usepackage{amssymb}
\usepackage{color}
\usepackage{url}

\newcommand{\slashPsub}{\slash\hspace{-0.5em}P}

\newcommand{\Nb}{\bar N}
\newcommand{\Fp}{F_\pi}

\newcommand{\mpi}{m_{\pi}}

\newcommand{\dslash}[1]{#1 \llap{/\kern-0.5pt}}
\newcommand{\Dslash}[1]{#1 \llap{/\kern+1.2pt}}
\newcommand{\DDslash}[1]{#1 \llap{/\kern+2.3pt}}
\newcommand{\dslashh}[1]{#1 \llap{/\kern+1pt}}

\newcommand{\bea}{\begin{eqnarray}}
\newcommand{\eea}{\end{eqnarray}}
\newcommand{\bma}{\begin{pmatrix}}
\newcommand{\ema}{\end{pmatrix}}

\begin{document}
\begin{titlepage}

\vspace{2.0cm}

\begin{center}
{\Large\bf
Parity violation in neutron capture on the proton: \\
\vspace{2mm}
determining the weak pion-nucleon coupling}
\vspace{1.3cm}

{\large \bf   J. de Vries$^{1}$, N. Li$^{1}$, Ulf-G. Mei{\ss}ner$^{1,2,3,4}$,\\ [0.2em]A. Nogga$^{1,2,3}$, E. Epelbaum$^{5}$,   N. Kaiser$^{6}$}

\vspace{0.5cm}

{\large
$^1$
{\it Institute for Advanced Simulation, Institut f\"ur Kernphysik,
and J\"ulich Center for Hadron Physics, Forschungszentrum J\"ulich,
D-52425 J\"ulich, Germany}}

\vspace{0.25cm}
{\large
$^2$
{\it JARA\,\,-\,\,Forces and Matter Experiments, Forschungszentrum J\"ulich,
D-52425 J\"ulich, Germany}}

\vspace{0.25cm}
{\large
$^3$
{\it JARA\,\,-\,\,High Performance Computing, Forschungszentrum J\"ulich,
D-52425 J\"ulich, Germany}}

\vspace{0.25cm}
{\large
$^4$
{\it Helmholtz-Institut f\"ur Strahlen- und Kernphysik and Bethe Center for
Theoretical Physics, Universit\"at Bonn, D-53115 Bonn, Germany}}

\vspace{0.25cm}
{\large 
$^5$ 
{\it Institut f\"ur Theoretische Physik II, Ruhr-Universit\"at Bochum, 44780 Bochum, Germany}}

\vspace{0.25cm}
{\large
$^6$
{\it Physik Department T39, Technische Universit\"at M\"unchen, D-85747 Garching, Germany}}

\end{center}

\vspace{1.5cm}

\begin{abstract}
We investigate the parity-violating analyzing power in neutron capture on the proton at thermal energies in the framework of chiral effective field theory. By combining this analysis with a previous analysis of parity violation in proton-proton scattering, we are able to extract the size of the weak pion-nucleon coupling constant. The uncertainty is significant and dominated by the experimental error which is expected to be reduced soon. 

\end{abstract}

\vfill
\end{titlepage}

Although parity violation (PV) induced by the weak interaction is well understood at the level of elementary quarks, its manifestation at the hadronic and nuclear level is not that clear. This holds particularly true for the strangeness-conserving part of the weak interaction which induces PV in hadronic and nuclear systems. The Standard Model predicts PV forces between nucleons. However, their forms and strengths are masked by the nonperturbative nature of QCD at low energies. Combined with the difficulty of doing experiments with sufficient accuracy to extract parity-violating signals, hadronic PV is one of the least tested parts of the Standard Model. 

The understanding of low-energy strong interactions has increased tremendously by the use of effective field theories (EFTs). It has been realized that by writing down the most general interactions among the low-energy degrees of freedom that are consistent with the symmetries of QCD, one obtains an EFT, chiral perturbation theory ($\chi$PT), that is a low-energy equivalent of QCD. Each interaction in the chiral Lagrangian comes with a coupling strength, or low-energy constant (LEC), which needs to be extracted from data or computed in lattice QCD. In contrast to low-energy QCD itself, $\chi$PT allows one to calculate observables in a perturbative framework with expansion parameter $p/\Lambda_\chi$, where $p$ is the momentum scale of the process and $\Lambda_\chi\sim 1$ GeV, the scale where the EFT breaks down. 
Although nuclear physics is intrinsically nonperturbative, the nucleon-nucleon ($N\!N$) potential can be calculated perturbatively within $\chi$PT. The resulting chiral potential is then iterated to all orders to calculate $N\!N$-scattering and bound state properties. This framework is usually called chiral nuclear EFT (for recent reviews, see Refs. \cite{Epelbaum:2008ga,Machleidt:2011zz}).  

The success of chiral EFT in parity-conserving (PC) nuclear physics has led to an analogous program in the PV sector \cite{Kaplan:1992vj, Kaplan:1998xi, Zhu, deVries2, Viviani}. One starts with the four-quark operators that are induced when the heavy weak gauge bosons are integrated out. The next step entails constructing a PV chiral Lagrangian which contains all interaction terms that transform under chiral symmetry in the same way as the underlying four-quark operators. From the resulting chiral Lagrangian one then calculates the PV $N\!N$ potential and electromagnetic current. In the final step the obtained PV potential and current are applied, in combination with the PC chiral potential and current, in calculations of nuclear processes. The PV LECs appearing in the PV chiral Lagrangian can be fitted to some data and other PV processes can then be predicted. 

Although this sounds like a good strategy, in practice this procedure is complicated by the lack of data on PV processes. So far, hadronic PV has only been measured in a handful of experiments (see Refs.~\cite{Holsteinreview, Schindler:2013yua} for recent reviews). The longitudinal analyzing power (LAP), which would be zero in the limit of no PV, has been measured for proton-proton scattering at three different energies \cite{Kistryn:1987tq, Eversheim:1991tg, Berdoz:2001nu}, for proton-alpha scattering only at a single energy \cite{Henneck:1982cc, Lang:1985jv}, and recently for the first time a preliminary result has been reported for radiative neutron capture on the proton $\vec np\to d \gamma$ at thermal energies \cite{Crawford}. Nonzero parity-violating signals have also been found in more complex systems, as exemplified by the radiative decay of the ${}^{19}$F nucleus \cite{Elsener:1984vp, Elsener:1987sx} and the anapole moment of the Cesium atom \cite{Wood:1997zq}. 

The first full chiral EFT analysis of PV nuclear forces has been done in Ref.~\cite{Kaplan:1992vj} where it has been concluded that at leading order (LO) only a single interaction term appears:
\begin{equation}\label{PoddLO}
\mathcal L_{\slashPsub } = \frac{h_\pi}{\sqrt{2}} \Nb (\vec \pi\times \vec \tau)^3 N\,\,\,,
\end{equation}
written in terms of the pion isospin-triplet $\vec \pi$, the nucleon isospin-doublet $N= (p,\,n)^T$, and the weak pion-nucleon coupling constant $h_\pi$. The leading order PV potential arising from one-pion exchange takes the form
 \begin{equation}
V_{\text{OPE}}
= - \frac{g_{A}h_\pi}{ 2\sqrt{2} F_\pi} i(\vec \tau_1\times \vec \tau_2)^3 \frac{(\vec \sigma_1+\vec \sigma_2)\cdot \vec k }{\mpi^2+\vec k^{\,2}},
\label{onepion}
\end{equation}
in terms of the momentum transfer $\vec k = \vec p - \vec p^{\,\prime}$, where 
$\vec p$ and $\vec p^{\,\prime}$ are the incoming and outgoing nucleon momenta in the center-of-mass frame, and $\vec\sigma_{1,2}$ and $\vec\tau_{1,2}$ the nucleon spin- and isospin-operators, respectively. $\Fp = 92.4$\,MeV denotes the
pion decay constant, $\mpi = 139.57$\,MeV the charged pion mass, and $g_A=1.29$ 
the nucleon axial-vector coupling constant taking into account the Goldberger-Treiman discrepancy, in order to represent the strong $\pi N\!N$-coupling. 

Considering that there are no other terms at leading order,  the one-pion exchange (OPE) potential can be expected to give the dominant contribution to PV in nuclear processes. 
Nevertheless, despite decades of experimental effort the existence of a long-range PV $N\!N$ force has not been confirmed. This indicates that $h_\pi$ could be smaller than expected from naive dimensional analysis which predicts $h_\pi \sim G_F F_\pi \Lambda_\chi \sim 10^{-6}$ (consistent with the often-used estimate $h_\pi = 4.6\cdot10^{-7}$ of Ref.~\cite{Desplanques:1979hn}), with $G_F \simeq 1.67 \cdot 10^{-5}\,$GeV$^{-2}$ the Fermi coupling constant. In fact, the isovector nature of the weak pion-nucleon coupling already gives a natural suppression of $\sin^2 \theta_w \sim 1/4$   \cite{Kaplan:1992vj, Phillips:2014kna}, while a large-$N_c$ analysis indicates that $h_\pi$ is even further suppressed \cite{ Phillips:2014kna, Kaiser1, Meissner1}. A first lattice QCD calculation gave $h_\pi \simeq 10^{-7}$ \cite{hpilatt}. Finally, the absence of a PV signal in the $\gamma$-ray emission from ${}^{18}$F leads to the bound $h_\pi \leq 1.3 \cdot 10^{-7}$ \cite{Adelberger:1983zz,Adelberger:1983zz2,Haxton:1981sf}. 

The evidence in favor of a small value of $h_\pi$ is not conclusive. Large-$N_c$ arguments can be misleading, especially for pionic interactions, while the lattice calculation did not include disconnected diagrams. The bound from ${}^{18}$F depends on nuclear structure calculations of a relatively complicated nucleus and, despite being a careful work, might suffer from uncontrolled uncertainties. Finally, the Cesium anapole moment prefers a much larger value $h_\pi \simeq 10^{-6}$ although the involved uncertainties are also larger \cite{Haxton:2001zq,Haxton:2001mi}.
 It seems that the only conclusive method of determining the size of $h_\pi$ is through a fit to experiments using simple few-body processes which are theoretically much better under control. Unfortunately, only a few PV signals have been measured so far in such few-body processes. In recent work we investigated the data on $\vec p p$ scattering in a chiral EFT framework \cite{deVries2,deVries}. The main goal of this paper is to combine this analysis with the recent data on PV in radiative neutron  capture on the proton $\vec n p \to d \gamma$ and extract a value of $h_\pi$. An analysis of PV in the inverse process $\vec \gamma d \rightarrow n p$ within pionless EFT has recently been performed in Ref.~ \cite{Vanasse:2014sva}.

Our task gets complicated by two things. First of all, the OPE potential in Eq.~\eqref{onepion} changes the total isospin and does not contribute to $\vec p p$ scattering. The three data points still carry information on the size of $h_\pi$ because the analyzing power does depend on $h_\pi$ through the two-pion-exchange (TPE) diagrams \cite{Zhu, deVries, Viviani}. The TPE diagrams appear at higher order in the chiral counting where additional contributions in the form of PV $N\!N$ contact terms appear as well \cite{Zhu, Girlanda:2008ts,Phillips:2008hn}.
Secondly, although the PV OPE potential does contribute to $\vec np\to d \gamma$ capture, if the coupling constant $h_\pi$ is really as small as suggested, formally higher-order corrections can become relevant and need to be taken into account. Again such corrections appear as $N\!N$ contact terms. We discuss these subleading terms in the PV potential and the current at a later stage.

The other ingredients required for the calculation of PV observables are the PC $N\!N$ potential and the PC and PV electromagnetic currents. As in Ref.~\cite{deVries}, we apply here the next-to-next-to-next-to-leading order (N${}^3$LO) chiral EFT potential obtained in Ref.~\cite{Epelbaum:2004fk} and we refer the reader to this paper for all further details. The N$^3$LO potential exists for several values of the cut-off needed to regularize the scattering equation. Here, we 
regularize the PV potential in the same way as the PC potential via 
\begin{equation}
V_{PV}(\vec p,\,\vec p^{\,\prime}) \rightarrow e^{-p^6/\Lambda^6 } V_{PV}(\vec p,\,\vec p^{\,\prime}) e^{-p^{\prime\,6}/\Lambda^6}\,\,\,,
\end{equation}
where three choices for $\Lambda =\{450,\, 550,\,600\}$ MeV are applied, see Ref.~\cite{deVries, Epelbaum:2004fk}. TPE diagrams are regularized with a spectral cut-off $\Lambda_S = \{500,\, 600,\,700\}$ MeV \cite{Epelbaum:2004fk}. In recent work \cite{Epelbaum:2014efa} an alternative regularization scheme (formulated in coordinate space) has been proposed which better preserves the long-range nature of pion-exchange terms in the potential. Considering the large experimental uncertainties in the field of nuclear parity violation, we do not expect 
drastic changes if the alternative regularization scheme is applied. Nevertheless, we will investigate this new scheme and its extension to N${}^4$LO \cite{Epelbaum:2014sza} in future work. 

Within the chiral EFT power-counting rules the dominant PC current arises from the nucleon magnetic moments. At next-to-leading (NLO) order we encounter the one-body convection current\footnote{Here the power-counting rules of Ref.~\cite{Epelbaum:2004fk} are followed where recoil and relativistic corrections are relegated to higher order by counting $1/m_N \sim k/\Lambda_\chi^2$, where $k$ is the typical momentum scale of the process. The magnetic moment operator is not a recoil correction and only scales as $1/m_N$ for conventional reasons. We thus treat $\mu_{s,v}/m_N \sim 1/\Lambda_\chi$ which is also justified by the large value of $\mu_v=4.72$.}, which arises from gauging the nucleon kinetic energy term, and the leading OPE two-body currents. The total PC current up to NLO is then given by
\begin{eqnarray}
\vec J_{PC} &=& \sum_{j=1}^2 \frac{e}{4m_N}\left\{ -\left[ \mu_s + \mu_v\tau_j^3 \right] i (\vec \sigma_j \times \vec q) + (1+\tau_j^3)(\vec P_j + \vec P_j^{\,\prime})\right\}\delta^{(3)}[\vec P_j -\vec P_j^{\,\prime}-\vec q\, ]\nonumber\\
&& +\frac{e g_A^2}{4\Fp^2} i\left(\vec \tau_1\times \vec \tau_2\right)^3
\left\{ 2\vec{k}\,
\frac{\vec \sigma_1\cdot(\vec k+\vec{q}/2)}{(\vec k +\vec q/2)^2+\mpi^2} \,
\frac{\vec \sigma_2\cdot(\vec k -\vec q/2)}{(\vec k -\vec q/2)^2+\mpi^2}  
\right.
\nonumber\\
&& \left. -\vec \sigma_1\,
\frac{\vec \sigma_2\cdot(\vec k-\vec q/2)}{(\vec k -\vec q/2)^2+\mpi^2}-
\vec \sigma_2\,
\frac{\vec \sigma_1\cdot(\vec k+\vec q/2)}{(\vec k +\vec q/2)^2+\mpi^2} 
\right\}\,\,\, , 
\end{eqnarray}
where $\mu_s = 0.88$  and $\mu_v=4.72$ are the isoscalar  and isovector nucleon magnetic moments. The momenta of the incoming and outgoing nucleon interacting with the photon (of outgoing momentum $\vec q\,$) are denoted by $\vec P_j$  and 
$\vec P^{\,\prime}_j$, respectively. The momenta carried by the  
intermediate pions are $\vec k+\vec q/2= \vec P_1-\vec P^{\,\prime}_1$ and 
$\vec k-\vec q/2= \vec P^{\,\prime}_2- \vec P_2$.  
In contrast, the leading PV current is solely due to OPE diagrams where one of 
the pion-nucleon vertices is from Eq.~\eqref{PoddLO} 

\begin{eqnarray}
\vec J_{PV} &=& \frac{eg_A h_\pi}{2\sqrt{2}\Fp}\left(\vec \tau_1\cdot\vec \tau_2 - \tau_1^3 \tau_2^3 \right) \bigg\{2 \vec k \frac{\vec \sigma_1\cdot(\vec k+\vec{q}/2)+\vec \sigma_2\cdot(\vec k-\vec{q}/2)}{[(\vec k +\vec q/2)^2+\mpi^2][(\vec k -\vec q/2)^2+\mpi^2]}
\nonumber\\
&& 
-\frac{\vec \sigma_1}{(\vec k -\vec q/2)^2+\mpi^2}-
\frac{\vec \sigma_2}{(\vec k +\vec q/2)^2+\mpi^2} 
\bigg\}\,\,\,.
\end{eqnarray}

These ingredients are sufficient to calculate the LO contribution to the 
longitudinal analyzing power in $\vec np \to d \gamma$. The details of the 
actual calculation will be presented in a longer paper~\cite{LongPV} and 
therefore we focus here just on the results. 

\begin{table}[t]
\caption{Total cross section in mb for unpolarized $np$ capture at $2.52\cdot 10^{-8}$ MeV lab energy. The first column is the calculated cross section using N${}^3$LO chiral potentials and the isovector nucleon magnetic moment $\mu_v$. The second column also includes the leading PC OPE current. The experimental result is from Ref. \cite{Cox}. Contributions from other currents at this order such as the isoscalar magnetic moment and convection current are negligible.}
\begin{center}\small
\begin{tabular}{||c|ccc||}
\hline
& isovector magnetic moment  & + PC OPE currents   & Experimental result\\
\hline
\rule{0pt}{3ex}
$\sigma_{\rm tot}$ & $305\pm 4$ & $ 319\pm 5 $  &$334.2 \pm 0.5$\\
 \hline
\end{tabular}
\end{center}
\label{tablecross}
\end{table}

The longitudinal analyzing power (LAP) in $\vec np \to d \gamma$ capture is defined as 
\bea\label{LAP}
A_\gamma(\theta) & = &\frac{d\sigma_+(\theta) - d \sigma_-(\theta)}{d\sigma_+(\theta) + d \sigma_-(\theta)} = a_\gamma\,\cos \theta\,\,\,,
\eea
with $d\sigma_\pm(\theta)$ the differential cross section for incoming neutrons with positive/negative helicity and $\theta$ the angle between the outgoing photon momentum $\vec q$ and the neutron spin. The experiment takes place at thermal energies where the total cross section for $np$ capture is dominated by the nucleon isovector magnetic moment $\mu_v$, while the isoscalar magnetic moment $\mu_s$ and the convection current give negligible contributions. At NLO the PC OPE currents add to the cross section at the $5\%$ level as can be seen in Table~\ref{tablecross}. For comparison, using the AV18 interaction we obtain $324$ mbar in good agreement with Refs.~\cite{Carlson:1997qn,Schiavilla:2002uc}. 
The remaining discrepancy of roughly $4\%$ with respect to the experimental result should be removed by higher-order corrections, for example in the form of PC contact and TPE currents. In phenomenological models indeed the remaining discrepancy is explained by heavy-meson-exchange currents \cite{Carlson:1997qn}. 
The theoretical uncertainties (1\%-2\%) quoted in the table are obtained from varying the cut-off parameters in the N${}^3$LO potential.

\begin{table}[t]
\caption{Contributions to the LAP $a_\gamma$ in $np$ capture in units of $h_\pi$. Part $1$ is the contribution from one-body currents only, Part $2$ from the isovector magnetic moment in combination with the PC OPE currents and the PV OPE potential, and Part $3$ from the interference of the isovector magnetic moment and the PV OPE currents.}
\begin{center}\small
\begin{tabular}{||c|cccc||}
\hline
& Part 1  & Part 2   & Part 3 & Total \\
\hline
\rule{0pt}{3ex}
$a_\gamma/h_\pi$ & $-0.27\pm0.03$ & $ -0.53\pm 0.02 $  &$0.72\pm 0.03$ & $-0.11\pm0.05$\\
 \hline
\end{tabular}
\end{center}
\label{tableLO}
\end{table}

Even in the presence of the PV potential, the numerator in Eq.~\eqref{LAP} vanishes if we only include the leading magnetic moment currents. An interference with electric dipole currents, which appear at NLO in the form of the convection and OPE currents, is necessary to obtain a non-vanishing result. The dominant contributions to $a_\gamma$ then consist of an interference between the isovector magnetic moment and: 
\begin{enumerate}
\item the one-body convection current in combination with the  PV OPE potential,
\item the two-body PC OPE currents in combination with the  PV OPE potential,
\item the two-body PV OPE currents.
\end{enumerate}
All these contributions appear at the same order in the chiral counting and we present the results in Table~\ref{tableLO}. The individual contributions are all of the same order, as expected from the power counting, and suffer only from minor uncertainties due to cut-off variations. However, the total result has a much larger relative uncertainty due to cancellations between the individual contributions. These cancellations were found also in Ref.~\cite{Hyun:2001yg} where the AV18 potential has been applied in combination with the same currents. Our central value is also in good agreement with results based on various phenomenological strong potentials and the Siegert theorem for the electric dipole currents \cite{Schiavilla:2002uc, Hyun:2004xp, Liu:2006dm}. These calculations do, however, 
not provide an uncertainty estimate. In Ref.~\cite{Hyun:2006cb} a smaller uncertainty was found when varying the cut-off (roughly $\pm 0.015\,h_\pi$), but the authors did not vary the strong potential simultaneously. In addition the Siegert theorem was applied for the electric dipole currents. The significant dependence of the total result on the cut-off parameters indicates that the extraction of $h_\pi$ from data on $a_\gamma$ is less  clean than might be expected. 

Having available the prediction for $a_\gamma$ as a function of $h_\pi$, we can now compare to data. For a long time only a bound on $a_\gamma$ existed
\begin{eqnarray}
a_\gamma = (0.6\pm2.1)\cdot10^{-7},\qquad
a_\gamma = (-1.2\pm1.9\pm0.2)\cdot 10^{-7}\,\,\,,
\end{eqnarray}
from Refs.~\cite{Cavaignac:1977uk} and \cite{Gericke:2011zz}, respectively. Applying our most conservative estimate we obtain an upper bound
$|h_\pi| \leq  4.5 \cdot 10^{-6}$ on the weak pion-nucleon coupling. Recently a first preliminary result for $a_\gamma$ was reported~\cite{Crawford} 
\begin{equation}\label{agammaEXP}
a_\gamma = (-7.14\pm4.4)\cdot 10^{-8}\,\,\,.
\end{equation}
This result is based on a subset of the full data taken and 
an improved result with an uncertainty at the $10^{-8}$ level is expected in the near future. 

 \begin{figure}[!t]
\centering
\includegraphics[width=0.49\textwidth]{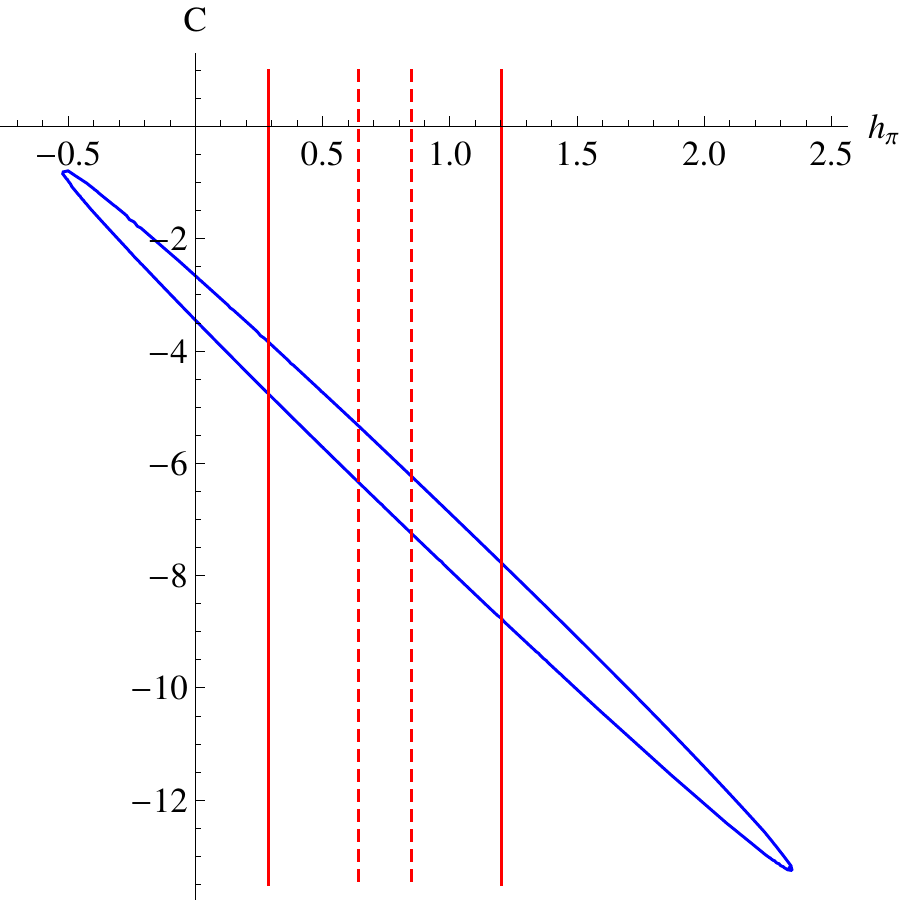} 
\includegraphics[width=0.49\textwidth]{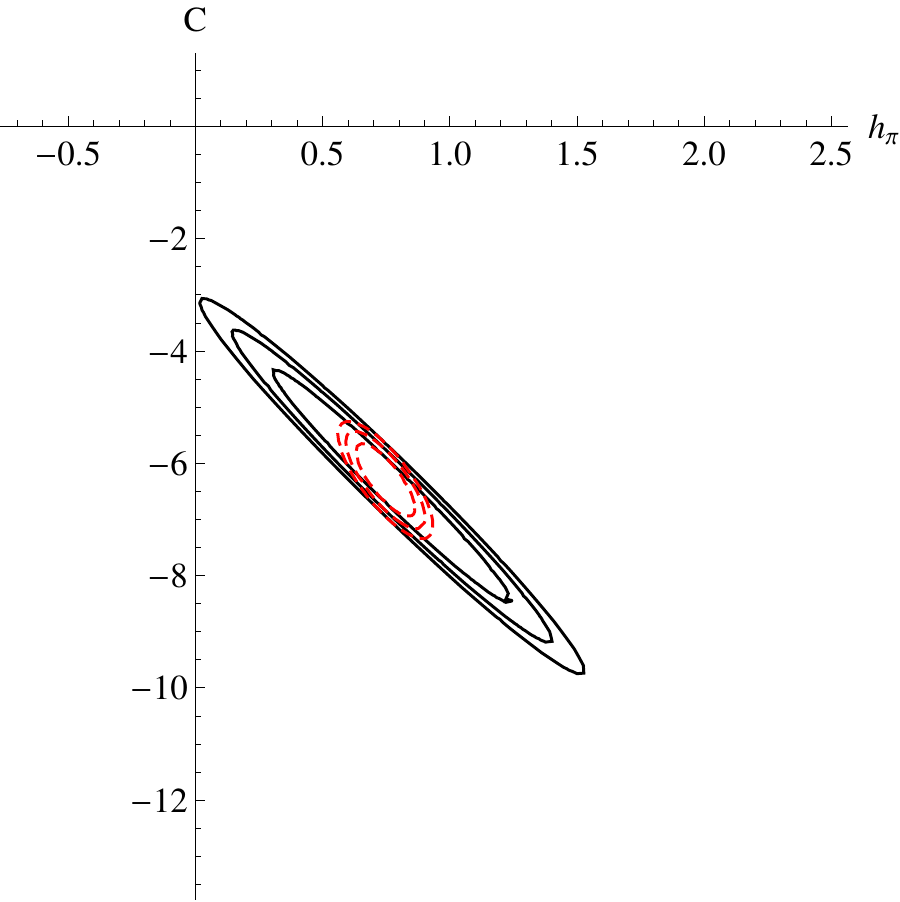}  
\caption{The allowed ranges for the LECs $h_\pi$ and $C$ (both in units 
of $10^{-6}$). Left panel: The blue ellipse is a fit to the $pp$ data with a total $\chi^2=2.71$ and the vertical solid lines the fit of $h_\pi$ to $a_\gamma$ at the one-sigma level. The vertical dashed lines correspond to the same fit, but now using the expected experimental uncertainty $(\pm 1\cdot 10^{-8})$ keeping the central value as in Eq.~\eqref{agammaEXP}. Right panel: The black (solid) ellipses are fits to the combined $pp$ and $np$ data with a total $\chi^2=2,3,4$. The red (dashed) ellipses are the same, but using the expected future experimental uncertainty.}
 \label{CUT3}
\end{figure}

To fit $h_\pi$ we combine the $a_\gamma$ analysis with that of the LAP in $pp$ scattering. As mentioned, the $pp$ LAP does not depend on the OPE PV potential in Eq.~\eqref{onepion} due to its isospin-changing nature. Nevertheless, the $pp$ LAP still depends on $h_\pi$ due to the TPE potential which appears at NLO. At the same order the PV potential contains PV $N\!N$ contact terms, but only one combination with LEC $C$ contributes to $pp$ scattering ($C = - C_0 +C_1 +C_2 - C_3$ in the notation of Ref.~\cite{deVries2}).
The LECs $h_\pi$ and $C$ were fitted to the $pp$ data in Refs.~\cite{deVries}. In Ref.~\cite{deVries2} the fit was slightly improved by including the dominant piece of the N${}^2$LO PV potential which does not depend on additional unknown LECs. 

We compare the obtained values of $h_\pi$ and $C$ from the data on the $np$ and $pp$ LAPs in Fig.~\ref{CUT3} using the intermediate cut-off values to regularize the potential. In the left panel, the ellipse denotes the contours of a total $\chi^2 =2.71$ corresponding to a fit to $pp$ data, whereas the red vertical lines denote a fit to $a_\gamma$ at the one-sigma level. In the right panel, the ellipses denote contours of a total $\chi^2=2,3,4$ corresponding to a combined fit to the $pp$ data and $a_\gamma$. The dashed lines in both panels are obtained if we use the expected future experimental uncertainty, $\pm 1 \cdot 10^{-8}$, of the measurement of $a_\gamma$ while using the same central value as in Eq.~\eqref{agammaEXP}. The dashed contours are only there to illustrate what the accuracy could be with better data but should not be used to extract values of the LECs.

 \begin{figure}[!t]
\centering
\includegraphics[width=0.49\textwidth]{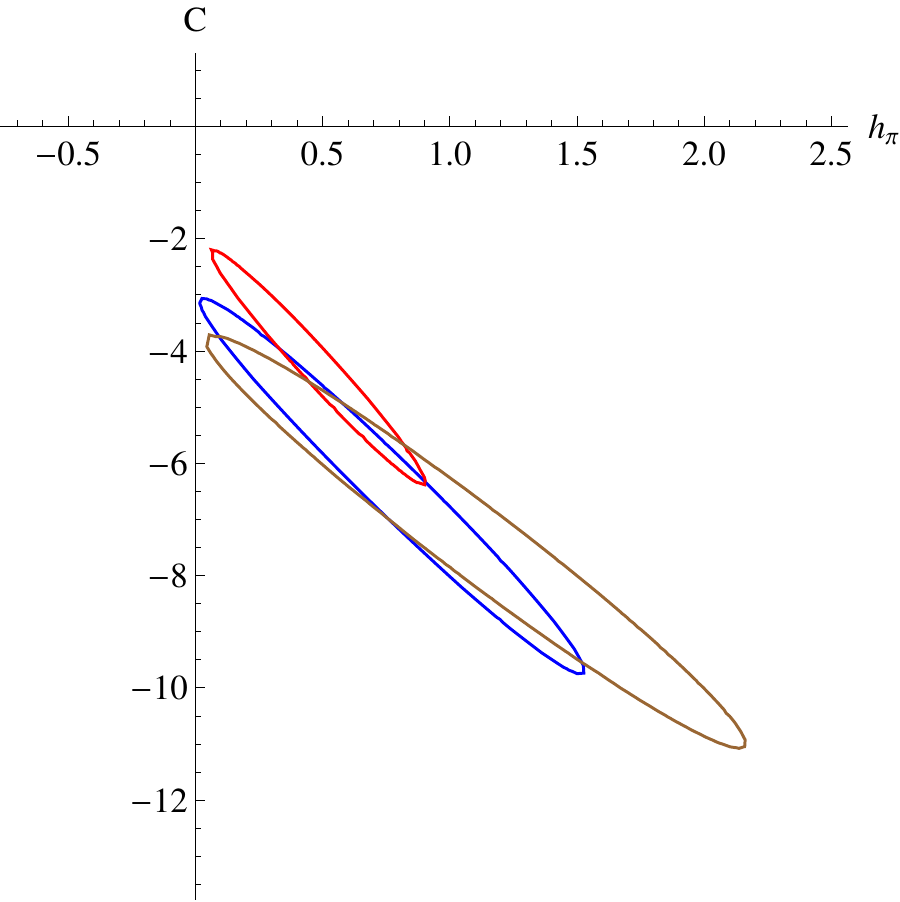}
\caption{The allowed ranges for the LECs $h_\pi$ and $C$ (both in units 
of $10^{-6}$). The ellipses correspond to fits to the combined $pp$ and $np$ data with a total $\chi^2=4$ for the three different cut-offs applied (from the smallest to the largest ellipse the cut-off changes from $450$ to $600$ MeV). 
}
 \label{CUT135}
\end{figure}
From the right panel of Fig.~\ref{CUT3} at the level of a total $\chi^2 =4$ the LECs $h_\pi$ and $C$ become
\begin{equation}
h_\pi = (0.80 \pm 0.70)\cdot 10^{-6}\,\,\,,\qquad C = (-6.0\pm 3.0)\cdot 10^{-6}\,\,\,.
\end{equation}
Since a $\chi^2$ analysis with so few data points can be misleading we collect in Table~\ref{tablefit} the observables for the $pp$ and $np$ systems using three different fit values that all lie within the contours of Fig.~\ref{CUT3}. From Table~\ref{tablefit} we see that small (large) values of $h_\pi = 10^{-7}$ ($h_\pi = 1.5\cdot 10^{-6}$) still give a reasonable fit to the $pp$ data, but underpredict (overpredict) $a_\gamma$. However, considering the large experimental uncertainty of $a_\gamma$ the fits cannot be excluded at a significant level.

\begin{table}[t]
\caption{Predictions for the $pp$ LAP $A_z$ and $a_\gamma$ (both in units of $10^{-7}$) for three fits using the intermediate cut-off combination. The first fit corresponds to the best-fit value with $h_\pi = 0.77$ and $C=-6.4$. The second and third fits correspond to the values at the edge of the contours with $h_\pi =0.1$ and $C=-3.0$, and $h_\pi=1.5$ and $C=-9.0$ respectively (all in units of $10^{-6}$). The first three columns correspond to the $pp$ LAP at three different energies and the fourth column to $a_\gamma$. The experimental results are from Refs. \cite{Kistryn:1987tq, Eversheim:1991tg, Berdoz:2001nu,Crawford}. }
\begin{center}\small
\begin{tabular}{||c|cccc||}
\hline
   &$A_z(13.6\, \mathrm{MeV})$ & $ A_z(45\, \mathrm{MeV}) $  &$A_z(221\, \mathrm{MeV})$ & $a_\gamma$ \\
\hline
\rule{0pt}{3ex}
Fit 1 & $-0.90$ & $-1.56$  &$0.57$ & $-0.74$ \\
Fit 2 & $-0.65$ & $-1.36$  &$0.50$ & $-0.10$ \\
Fit 3 & $-0.89$ & $-1.19$  &$0.43$ & $-1.44$ \\
Exp. & $-0.93\pm0.21$ & $-1.50\pm0.22$ & $0.84\pm0.34$ & $-0.71\pm 0.44$\\
 \hline
\end{tabular}
\end{center}
\label{tablefit}
\end{table}

To study the dependence of the extraction of the LECs on the details of the strong $N\!N$ potential, we repeat the analysis for other cut-off values. For $\Lambda =\{450,\, 550,\, 600\}$ MeV, respectively, the following best fit parameters emerge
\begin{eqnarray}
h_\pi &=& \{0.48,\, 0.77,\, 1.1\}\cdot 10^{-6}\,\,\,,\nonumber\\
C   &=& -\{4.3,\, 6.4,\, 7.4  \}\cdot 10^{-6}\,\,\,.
\end{eqnarray}
We see that the uncertainty in $h_\pi$ due to cut-off variations is roughly $40\%$. At present, the experimental uncertainty is still larger. However, once the precision of the $a_\gamma$ measurement is improved by roughly a factor $2$ the cut-off dependence will dominate the uncertainty. 
The observed cutoff dependence in the predicted value of
$a_\gamma$ is likely to be considerably reduced by the inclusion of
higher-order corrections to the exchange currents. One frequently used
approach along this line is to make use of the Siegert theorem. We
emphasize, however, that such an approach yields only incomplete
results for the exchange currents. We postpone a detailed investigation of the
role played by higher-order contributions to a future study. 

In Fig.~\ref{CUT135} we show contours for a total $\chi^2=4$ for three different cut-off values. To obtain an allowed range of the LECs, we extract the minimal and maximal values allowed by the three contours. With this conservative approach we obtain the following ranges
\begin{equation}
h_\pi = (1.1 \pm 1.0)\cdot 10^{-6}\,\,\,,\qquad C = (-6.5\pm 4.5)\cdot 10^{-6}\,\,\,.
\end{equation}
The fits indicate that small values of $h_\pi \sim 10^{-7}$ are barely consistent with the data, with values of $h_\pi\sim(5-10)\cdot 10^{-7}$ being preferred.  Such larger values disagree with the upper limit from ${}^{18}$F gamma-ray emission $h_\pi \leq 1.3\cdot 10^{-7}$ and lattice and model calculations of $h_\pi \simeq 10^{-7}$. An increase in the accuracy of the $a_\gamma$ measurement is needed to make a firmer statement about this discrepancy.

So far, our analysis included only the LO contribution to $a_\gamma$ proportional to $h_\pi$. If for whatever reason $h_\pi$ is small, formally subleading contributions might actually be dominant. The first corrections to $a_\gamma$ appear two orders down in the chiral expansion and in principle consist of two contributions. The first arises from TPE diagrams in both the PV potential and currents. However, these contributions are proportional to $h_\pi$ as well and are then additionally suppressed by the assumed smallness of $h_\pi$ (in Ref.~\cite{Hyun:2006cb} TPE contributions were found at the 10\% level with respect to the OPE result based on the Siegert theorem and a phenomenological $N\!N$ model). The other corrections appear in the form of PV $N\!N$ interactions which contribute both to the PV potential and current. It turns out that the relevant potential and current depend on the same LEC $C_4$, in the notation of Ref.~\cite{deVries2},  which is independent of the LEC $C$ appearing in $pp$ scattering. 
\begin{eqnarray}\label{NLO}
V_{\text{PV,NLO}}&=& \frac{C_4}{ \Fp \Lambda_\chi^2} i(\vec \tau_1\times \vec \tau_2)^3(\vec \sigma_1 + \vec \sigma_2)\cdot \vec k,\\
\vec J_{\text{PV,NLO}}&=&- \frac{C_4}{\Fp \Lambda_\chi^2}\left(\vec \tau_1\cdot\vec \tau_2 - \tau_1^3 \tau_2^3 \right) (\vec \sigma_1+\vec \sigma_2)\,\,\,.
\end{eqnarray}
Other contact contributions to the PV potential and current appearing at this order give rise to negligible contributions to $a_\gamma$. 

We obtain the total result for the asymmetry
\begin{equation}\label{total}
a_\gamma = (-0.11\pm0.05) h_\pi + (0.055\pm0.025) C_4\,\,\,.
\end{equation}
To estimate the size of the $C_4$ contributions to $a_\gamma$ we can use resonance saturation. By comparison with the meson-exchange model of Ref.~\cite{Desplanques:1979hn}, usually called the DDH model, the LEC $C_4$ can be expressed as\footnote{In Ref.~\cite{deVries2} the resonance-saturation estimate of $C_4$ was found to also depend on $h_\pi$ due to TPE diagrams. However, in the calculation of $a_\gamma$ we do not include TPE contributions explicitly so these terms should not be subtracted from the estimate.}
\begin{equation}
C_4 = \frac{F_\pi \Lambda_\chi^2}{2 m_N} \left[\frac{g_\omega h_\omega^1}{m_\omega^2} + \frac{g_\rho (h^{1\,\prime}_\rho-h^1_\rho)}{m_\rho^2} \right] \,\,\,,
\end{equation}
in terms of the masses $m_\rho\simeq m_\omega \simeq 780$ MeV and PC couplings $g_\omega = 8.4$ and $g_\rho = 2.8 $. We have checked that both the potential and current in Eq.~\eqref{NLO} \cite{LongPV} depend on this combination of DDH parameters by comparing to the currents derived in Ref.~\cite{Haxton:2001zq}. 
The sizes of the PV couplings $h_\omega^1$, $h_\rho^1$, and $h^{1\,\prime}_\rho$ are unknown but can be estimated, albeit with significant uncertainty. Taking into account the whole reasonable range for these couplings as obtained in Ref.~\cite{Desplanques:1979hn}, we find $C_4 = (-0.8\pm0.4)\cdot 10^{-7}$. This range includes the more accurate prediction $C_4 = -1.2\cdot 10^{-7}$ of Refs.~\cite{Kaiser1, Meissner1}. To be conservative, we insert the DDH range into 
Eq.~\eqref{total} to obtain the estimated uncertainty due to the short-range PV $N\!N$ interaction
\begin{equation}
a_\gamma = (-0.11\pm0.05) h_\pi - (0.5\pm0.5)\cdot 10^{-8}\,\,\,.
\end{equation}
Considering the current experimental uncertainty of $\pm 4.4 \cdot 10^{-8}$, the contact terms provide only a minor error. This would imply that the above analysis and extraction of $h_\pi$ is reliable. Small values of $h_\pi$ are thus disfavored, although formally not (yet) inconsistent. An improvement in the measurement of $a_\gamma$ will provide a more definite answer regarding the size of $h_\pi$. In Refs.~\cite{Schiavilla:2002uc, Liu:2006dm, Gericke:2011zz} the dependence of the asymmetry on the short-range DDH parameters is found to be smaller than the central value of Eq. \eqref{total} by roughly a factor $4$ to $5$. A possible explanation might be the use of  phenomenological strong potentials that typically have a stronger short-range repulsion than the chiral EFT potential, leading to a smaller dependence on short-range operators. A similar effect was found in the study of electric dipole moments \cite{Bsaisou:2014zwa}.

 \begin{figure}[!t]
\centering
\includegraphics[width=0.49\textwidth]{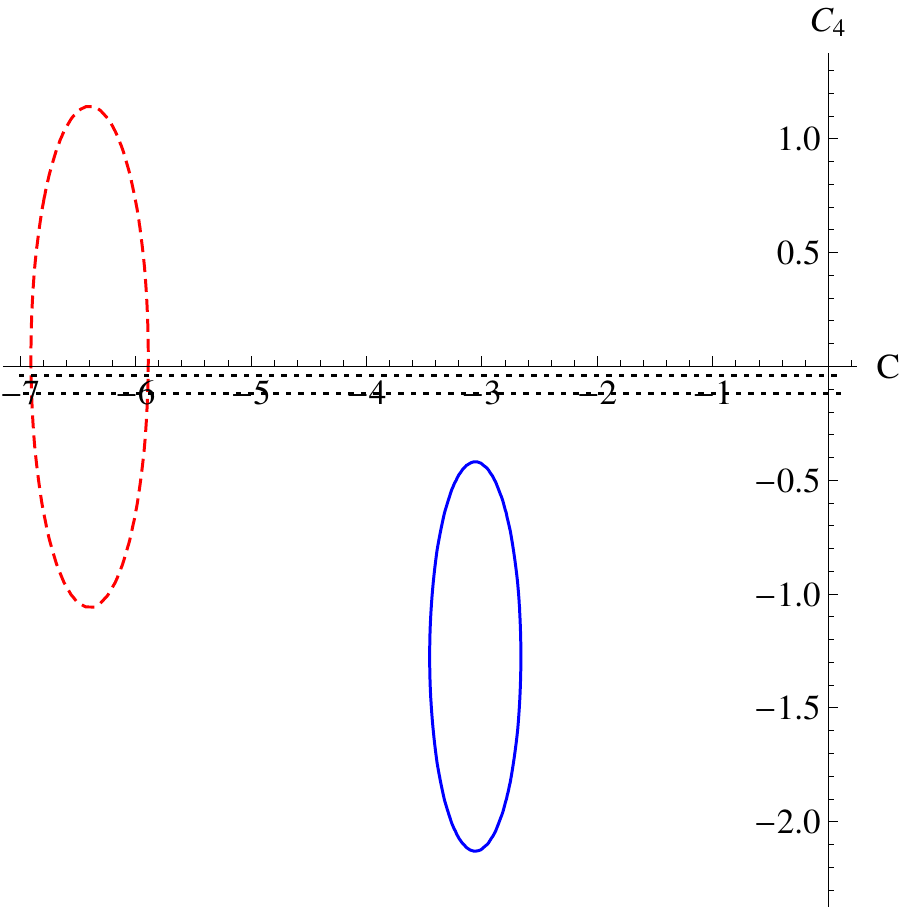}  
\caption{The allowed ranges for the LECs $C$ and $C_4$ (both in units 
of $10^{-6}$). The blue (solid) ellipse is a fit to the $pp$ and $np$ data with a total $\chi^2=2.71$ using $h_\pi=0$. The red (dashed) ellipse is similar but now using $h_\pi =7.7\cdot 10^{-7}$ which is the best-fit value for the intermediate cut-off. The dotted horizontal lines mark the predictions made by resonance saturation.}
 \label{CCg}
\end{figure}

The above reasoning is to some extent circular. We want to fit the LECs from few-body data only, but to estimate the effects of the formally subleading correction we require a model estimate of $C_4$. This unfortunate situation is due to a lack of data which implies that we cannot fit all LECs at the same time in a consistent way. Some more insight can be obtained by using an alternative strategy. We force $h_\pi$ to be small and fit the LECs $C$ and $C_4$ to the $pp$ and $np$ data. We then obtain the fits in Fig.~\ref{CCg}. The blue (solid) contour corresponds to a fit in the $C-C_4$ plane with a total $\chi^2=2.71$ where we set $h_\pi=0$. In this case the fit prefers values for $C_4$ which lie outside the range obtained from resonance saturation. The red (dashed) contour corresponds to a fit using $h_\pi = 7.7 \cdot 10^{-7}$  which corresponds to the best fit value for the intermediate cut-off combination. In this case the fit for $C_4$ is centered around the zero and includes the resonance-saturation range marked by the dotted lines. Although strong conclusion cannot be drawn from this observation, it does indicate that small values of $h_\pi$ requires short-range contributions that are larger than expected. As always, more and/or more precise data are required to draw firmer conclusions. 

To summarize, in this paper we have extracted the values of two low-energy constants $h_\pi$ and $C$ appearing in the parity-violating nucleon-nucleon potential and currents. To do so, we have used data on parity violation in proton-proton scattering and radiative neutron capture on a proton target.  The extraction has been performed in the framework of chiral effective field theory which has been systematically applied to both the parity-conserving and parity-violating parts of the problem. We have estimated the uncertainties of the fits due to experimental uncertainties, variation of cut-off parameters, and higher-order corrections and find the first of these to be dominant. Our extraction of the weak pion-nucleon coupling constant, $h_\pi = (1.1\pm 1.0)\cdot 10^{-6}$, is marginally consistent with bounds obtained in experiments on ${}^{18} F$ and a lattice QCD calculation. 
The expected increase in sensitivity of the $a_\gamma$ measurement will significantly improve the fit and tell whether small values of $h_\pi$ are consistent with few-body experiments.

\subsection*{Acknowledgements}
We thank Christopher Crawford for clarifications regarding the preliminary result of the asymmetry.
This work is supported in part by the DFG and the NSFC
through funds provided to the Sino-German CRC 110 ``Symmetries and
the Emergence of Structure in QCD'' (Grant No. 11261130311).
We acknowledge the support of the European Community-Research Infrastructure 
Integrating Activity ``Study of Strongly Interacting Matter'' (acronym
HadronPhysics3, Grant Agreement n. 283286) under the Seventh Framework Programme of EU. Parts of the calculations presented 
here were performed at the J\"ulich Supercomputing Center, J\"ulich, Germany.

\end{document}